\documentclass[reprint,showpacs,preprintnumbers, API,prb,]{revtex4-1}
\usepackage{graphicx}
\usepackage{dcolumn}
\usepackage{bm}

\begin{document}

\title{Chiral Polaron Formation in Graphene}
\author{B. S. Kandemir}
\date{\today}

\begin{abstract}
A theoretical investigation of the possible existence of the chiral polaron
formation in graphene is reported. We present an analytical method to
calculate the ground-state of the electron-phonon system within the
framework of the Lee-Low and Pines theory. On the basis of our model, the
influence of electron-optical phonon interaction onto the graphene
electronic spectrum is examined. In this paper, we only considered doubly
degenerate optical phonon modes of $\mathrm{E}_{2g}$ symmetry near the zone
center $\Gamma $. We show analytically that the energy dispersions of both
valance and conduction bands of the pristine graphene differ significantly
than those obtained through the standard electron self energy calculations
due to the electron-phonon interactions. Furthermore, we prove that the
degenerate band structure of the graphene promote the chiral polaron
formation.
\end{abstract}

\maketitle

\affiliation{Department of Physics, Faculty of Sciences, Ankara University, 06100\\
Tando\u{g}an, Ankara, Turkey}

Since the discovery of graphene \cite{Novoselov2004}, many studies have
focused on electron-phonon interactions, because of their particular
importance in understanding electronic and optical properties of graphene
\cite%
{Lazzeri2005,Lazzeri2006,Pisana2007,Yan2007,Calandra2007,Basko2008,Stauber2008a, Vladimir2010,Goerbig2007,Basko2007,Basko2008r,Lazzeri2008,Mariani2008,Faugeras2009, Carbotte2010,Mariani2010,Hwang2010,Ping2012,Araujo2012}%
. Even though the interaction of electron with doubly degenerate optical
phonon modes of  $\mathrm{E}_{2g}$ symmetry near the zone center $\Gamma $
does not open a gap \cite{Pisana2007,Dubay2003}, they contribute
significantly to intravalley-intraband and intravalley-interband scattering
\cite{Rana2009}. Since these phonon modes consist of in-plane out-of-phase
displacements of two sublattices A and B, they just shift the $\mathrm{K}$ ($%
\mathrm{K}^{^{\prime }}$) point so as to electronic bands are still
described by Dirac cones \cite{Pisana2007}. Moreover, it is also revealed
that the gate-modulated low-temperature Raman spectroscopy, the graphene G
band, which is the optical phonon at long wavelength, is markedly sensitive
to the coupling with Dirac fermion excitations at small wave vectors (long
wavelengths) \cite{Yan2007}.

So far only perturbational methods for intraband transitions have been
applied to the electron-phonon interactions, the combined effect of intra-
and inter-band scattering in a single valley has not been considered. Due to
the fact that graphene is a semimetal, actually gapless or zero-gap
semiconductor, the Dirac cones can not be treated independently for electron-%
$\mathrm{E}_{2g}$ phonon interactions. Single band approximations take into
account the intraband transitions only and they ignore the lack of gap, thus
omit the electron (hole) transitions from $\pi $ $(\pi ^{\ast })$ valance
(conduction) to $\pi ^{\ast }$($\pi $ ) conduction (valance) bands. Whereas,
due to the lack of gap, $\mathrm{E}_{2g}$ phonon can easily excite
electron-hole pairs leading to the chirality dependent modifications in
carrier dynamics.

In this letter, we used the continuum Fr\"{o}hlich type model to treat the
interaction of electron with long wavelength $\mathrm{E}_{2g}$ phonon. We
introduce a diagonalization procedure based on Lee-Low-Pines (LLP) like
transformations to investigate the properties of both valance and conduction
band polarons in graphene. Within the framework of low-energy continuum
model of graphene, the Hamiltonian of an electron interacting with optical $%
\Gamma $-phonon around the $\mathbf{K}$ point in the Brillouin zone can be
written as

\begin{equation}
\mathcal{H}=\mathcal{H}_{0}+\sum_{\boldsymbol{q}}\hbar \omega _{\mu }(%
\boldsymbol{q})b_{\boldsymbol{q\mu }}^{\dag }b_{\boldsymbol{q\mu }}+\mathcal{%
H}_{e-p},  \label{1}
\end{equation}%
where $\mathcal{H}_{0}=v_{F}\boldsymbol{\sigma }\cdot \boldsymbol{p}$ is the
unperturbed bare Hamiltonian, whose spectrum describes cone like behavior
with eigenvalues $\epsilon _{k\lambda }=\lambda \epsilon _{k}$ wherein $%
\epsilon _{k}=v_{F}k$ and $\lambda $ is the chirality index, together with
the corresponding eigenkets $\left\vert \boldsymbol{k}\lambda \right\rangle
=\exp (i\boldsymbol{k}\cdot \boldsymbol{r})\left(
\begin{array}{cc}
0 & \lambda e^{i\theta \left( \boldsymbol{k}\right) }%
\end{array}%
\right) ^{\dag }/\sqrt{2}L.$

In Eq.(\ref{1}), $\mathcal{H}_{e-p}$ is the electron-optical phonon
interaction Hamiltonian \cite{Basko2008,Lazzeri2008,Ando2006b}, and is given
by

\begin{equation}
\mathcal{H}_{e-p}=-\sum_{\boldsymbol{q}}\left[ \widetilde{\mathrm{M}}_{\mu
}^{\mathbf{K}}\left( \boldsymbol{q}\right) b_{\boldsymbol{q\mu }}e^{i%
\boldsymbol{q\cdot r}}+\mathrm{H.c.}\right] ,  \label{2}
\end{equation}%
where $b_{\boldsymbol{q\mu }}^{\dag }\left( b_{\boldsymbol{q\mu }}\right) $
is the optical phonon creation (annihilation) operator with longitudinal and
transverse optical phonon branch index $\mu =1$ (LO) and $2$ (TO),
respectively. Their dispersion have the form $\omega _{\mu }\left(
\boldsymbol{q}\right) =\omega _{\Gamma }\left( 0\right) R_{\mu }\left(
qa\right) $ with dimensionless part $R_{\mu }\left( qa\right) =\ \left[
1-r_{\mu }\left( qa\right) \right] ^{1/2}$, and with $\omega _{\Gamma
}\left( 0\right) =0.196$ $eV$, where $r_{1}$ and $r_{2}$ are given by $0.1138
$ and $0.01979$, respectively\cite{Stauber2008b}. The momentum dependent
matrix element of the interaction Eq.(\ref{2}) is $\widetilde{\mathrm{M}}%
_{\mu }^{\mathbf{K}}\left( \boldsymbol{q}\right) =\overline{\mathrm{M}}_{\mu
}^{\mathbf{K}}\left( \boldsymbol{q}\right) \mathrm{M}_{\mu }\left(
\boldsymbol{q}\right) $ , and defined as

\begin{equation}
\mathrm{M}_{\mu }\left( \boldsymbol{q}\right) =\left[
\begin{array}{cc}
0 & \mathrm{M}_{\mathrm{BA}}e^{-i\phi \left( \boldsymbol{q}\right) } \\
\mathrm{M}_{\mathrm{AB}}e^{i\phi \left( \boldsymbol{q}\right) } & 0%
\end{array}%
\right]  \label{3}
\end{equation}%
with $\mathrm{M}_{\mathrm{AB}}=+1(i)$ and $\mathrm{M}_{\mathrm{BA}}=-1(i)$ \
for LO(TO) phonons\cite{Tse2007}. $\phi \left( \boldsymbol{q}\right) =\tan
^{-1}\left( q_{y}/q_{x}\right) $ is the azimuthal angle of the phonon wave
vector $\boldsymbol{q,}$ and $\overline{\mathrm{M}}_{\mu }^{\mathbf{K}%
}\left( \boldsymbol{q}\right) =\overline{\mathrm{M}}/\sqrt{NR_{\mu }\left(
qa\right) }$ with $\overline{\mathrm{M}}=3a_{0}q_{0}J_{0}/\sqrt{2}$. Here, $%
q_{0}=\left( \partial J_{0}/\partial a\right) /J_{0}$ is predicted\cite%
{Pietronero1980,Jishi1993} around $2$ $\mathring{A}^{-1}$ or $2.5$ $%
\mathring{A}^{-1}$.

To solve Eq.(\ref{1}), we propose a diagonalization procedure based on the
LLP method, which includes two successive unitary transformations. Pristine
graphene has a electron-hole degeneracy point at $k=0$. Therefore, to be
compatible with this gapless band structure of the graphene, we make an
ansatz for the chiral polaron ground-state vector
\begin{equation}
\left\vert \boldsymbol{\Phi }\right\rangle _{pol}=U_{1}U_{2}\left\vert
\boldsymbol{0}\right\rangle _{ph}\otimes \left\vert \boldsymbol{\pm }%
\right\rangle   \label{4}
\end{equation}%
such that $\mathcal{H}\left\vert \boldsymbol{\Phi }\right\rangle
_{pol}=E_{\pm }\left\vert \boldsymbol{\Phi }\right\rangle _{pol}$. In Eq.(%
\ref{4}), $\left\vert \boldsymbol{0}\right\rangle _{ph}$ stands for the
phonon vacuum, and $\left\vert \boldsymbol{\pm }\right\rangle $=$%
\sum_{\lambda }\alpha _{\pm }^{\lambda }\left\vert \boldsymbol{k}\lambda
\right\rangle $ corresponds to electronic state vector defined through the
appropriate fractional amplitudes, $\alpha _{\pm }^{\lambda }$, due to the
fact that polaronic wavefunction must be the linear combination of $%
\left\vert \boldsymbol{k}+\right\rangle $ and $\left\vert \boldsymbol{k}%
-\right\rangle $, respectively. While the first unitary transformation
\[
U_{1}=\exp \left[ -i\boldsymbol{r\cdot }\sum_{\boldsymbol{q}}\boldsymbol{q}%
b_{\boldsymbol{q\mu }}^{\dag }b_{\boldsymbol{q\mu }}\right]
\]%
eliminates the electron coordinates, since the transformed operators are
given by the relations, $\widetilde{b}_{\boldsymbol{q\mu }}=b_{\boldsymbol{%
q\mu }}\exp \left[ -i\boldsymbol{q\cdot r}\right] $ and $\widetilde{%
\boldsymbol{p}}\boldsymbol{=p}-\sum_{\boldsymbol{q}}\hbar \boldsymbol{q}b_{%
\boldsymbol{q\mu }}^{\dag }b_{\boldsymbol{q\mu }}$, second unitary
transformation
\[
U_{2}=\exp \left\{ \sum_{\boldsymbol{q}}\left[ \overline{\mathrm{M}}_{0\mu
}^{\ast }\left( \boldsymbol{q}\right) \left\langle \boldsymbol{k}\lambda
^{^{\prime }}\right\vert \mathrm{M}_{\mu }^{\dag }\left( \boldsymbol{q}%
\right) \left\vert \boldsymbol{k}\lambda ^{{}}\right\rangle b_{\boldsymbol{%
q\mu }}^{\dag }-\mathrm{H.c.}\right] \right\}
\]%
is the displaced oscillator transformation with amplitude $\overline{\mathrm{%
M}}_{0\mu }\left( \boldsymbol{q}\right) =\overline{\mathrm{M}}_{\mu }^{%
\mathbf{K}}\left( \boldsymbol{q}\right) $/$\hbar \omega _{\mu }(\boldsymbol{q%
})$. It just shifts the phonon coordinates. As a result, the transformed
Hamiltonian can be written as $\widetilde{\mathcal{H}}=$ $%
U_{2}^{-1}U_{1}^{-1}\mathcal{H}U_{1}U_{2}.$ Therefore, $_{ph}\left\langle
\boldsymbol{0}\right\vert \widetilde{\mathcal{H}}\left\vert \boldsymbol{0}%
\right\rangle _{ph}$ leads to the following equation:
\begin{widetext}
\begin{eqnarray*}
_{ph}\left\langle \boldsymbol{0}\right\vert \widetilde{\mathcal{H}}%
\left\vert \boldsymbol{0}\right\rangle _{ph} &=&v_{F}\boldsymbol{\sigma }%
\cdot \left( \boldsymbol{p-}\hbar \sum_{\boldsymbol{q}}\boldsymbol{q}%
\left\vert \overline{\mathrm{M}}_{0\mu }\left( \boldsymbol{q}\right)
\right\vert ^{2}\left\vert \left\langle \boldsymbol{k}\lambda ^{^{\prime
}}\right\vert \mathrm{M}_{\mu }^{\dag }\left( \boldsymbol{q}\right)
\left\vert \boldsymbol{k}\lambda ^{{}}\right\rangle \right\vert ^{2}\right)
\\
&&+\sum\limits_{\boldsymbol{q}}\hbar \omega _{\mu }(\boldsymbol{q}%
)\left\vert \overline{\mathrm{M}}_{0\mu }\left( \boldsymbol{q}\right)
\right\vert ^{2}\left\vert \left\langle \boldsymbol{k}\lambda ^{^{\prime
}}\right\vert \mathrm{M}_{\mu }^{\dag }\left( \boldsymbol{q}\right)
\left\vert \boldsymbol{k}\lambda ^{{}}\right\rangle \right\vert
^{2}-\sum\limits_{\boldsymbol{q}}\left[ \mathrm{M}_{\mu }\left( \boldsymbol{q%
}\right) \left\vert \overline{\mathrm{M}}_{0\mu }\left( \boldsymbol{q}%
\right) \right\vert ^{2}\left\vert \left\langle \boldsymbol{k}\lambda
^{^{\prime }}\right\vert \mathrm{M}_{\mu }^{\dag }\left( \boldsymbol{q}%
\right) \left\vert \boldsymbol{k}\lambda ^{{}}\right\rangle \right\vert ^{2}+%
\mathrm{H.c.}\right] .
\end{eqnarray*}%
\end{widetext}Finally, by using the ansatz given by Eq.(\ref{4}), one can
easily construct characteristic equation of the matrix in the form
\begin{widetext}
\begin{eqnarray}
\left\vert
\begin{array}{cc}
E_{\pm }-\hbar v_{F}k\boldsymbol{+}\hbar v_{F}\boldsymbol{\Sigma }_{++}(\mu
,k)+\boldsymbol{\Sigma }_{++}^{0}(\mu ,k) & \hbar v_{F}\boldsymbol{\Sigma }%
_{+-}(\mu ,k)+2\boldsymbol{\Sigma }_{+-}^{0}(\mu ,k) \\
\hbar v_{F}\boldsymbol{\Sigma }_{-+}(\mu ,k)+2\boldsymbol{\Sigma }%
_{-+}^{0}(\mu ,k) & E_{\pm }+\hbar v_{F}k\boldsymbol{+}\hbar v_{F}%
\boldsymbol{\Sigma }_{--}(\mu ,k)+\boldsymbol{\Sigma }_{--}^{0}(\mu ,k)%
\end{array}%
\right\vert =0 \label{5}
\end{eqnarray}
\end{widetext}with elements

\begin{eqnarray}
\left.
\begin{array}{c}
\boldsymbol{\Sigma }_{++}(\mu ,k) \\
\boldsymbol{\Sigma }_{+-}(\mu ,k)%
\end{array}%
\right\}  &=&\frac{1}{2}\sum\limits_{\boldsymbol{q}}\left\vert \overline{%
\mathrm{M}}_{0\mu }\left( \boldsymbol{q}\right) \right\vert ^{2}  \nonumber
\\
&\times &\left[ 1\pm s_{\mu }\cos 2\left( \theta -\phi \right) \right]
\left\{
\begin{array}{c}
\boldsymbol{q}_{\mu }\cos \left( \theta -\phi \right)  \\
\boldsymbol{iq}\sin \left( \theta -\phi \right)
\end{array}%
\right. ,  \nonumber \\
\boldsymbol{\Sigma }_{\lambda \lambda ^{^{\prime }}}^{0}(\mu ,k) &=&\frac{1}{%
2}\sum\limits_{\boldsymbol{q}}\left\vert \overline{\mathrm{M}}_{0\mu }\left(
\boldsymbol{q}\right) \right\vert ^{2}  \nonumber \\
&\times &\hbar \omega _{\mu }(\boldsymbol{q})\left[ 1+s_{\mu }\lambda
\lambda ^{^{\prime }}\cos 2\left( \theta -\phi \right) \right] ,  \label{6}
\end{eqnarray}%
where $\boldsymbol{\Sigma }_{--}=-\boldsymbol{\Sigma }_{++}$ and $%
\boldsymbol{\Sigma }_{-+}=\boldsymbol{\Sigma }_{+-}^{\ast }$. \ After
converting the sums in Eq.(\ref{6}) into integrals over $\boldsymbol{q,}$ it
is easy to see that the terms with $\boldsymbol{\Sigma }_{\lambda \lambda
^{^{\prime }}}(\mu ,k)$ do not give contribution to the eigenvalue
calculation, since they all vanish after the $\boldsymbol{q}$ integration.
But the rest, i.e., terms with $\boldsymbol{\Sigma }_{\lambda \lambda
^{^{\prime }}}^{0}(\mu ,k)$ contribute. Thus, from the secular equation,
i.e., from Eq.(\ref{5}) the eigenvalues $E_{\pm }$ can then be solved
analytically as a function of $k$ in closed form,$\qquad $

\begin{equation}
E_{\pm }=\pm \left\{ \left( \hbar v_{F}k\right) ^{2}+4\left[ \boldsymbol{%
\Sigma }_{\pm \mp }^{0}(\mu ,k)\right] ^{2}\right\} ^{1/2}-\boldsymbol{%
\Sigma }_{\pm \pm }^{0}(\mu ,k).  \label{7}
\end{equation}%
In Eq.(\ref{7}), while the last term is due to the intraband transitions,
second one in the parenthesis is contribution due to the coupling between
valance and conduction bands, i.e. it corresponds to interband transitions.
Thus, our chiral polaron dispersion consists of partial mixture of
contributions from valance and conduction bands. While taking the integrals
in Eq.(\ref{7}), since they diverge at  upper limit of the integrations, we
must introduce an upper cut-off frequency in the integrations
to make them to be finite. It is nothing but just $2k$, due to
$q\in \left[ 0,2k \right]$. As a result,one obtains $k-$dependent contributions as
\begin{figure}[tbp]
{} \includegraphics[height=11.27cm,width=7.cm]{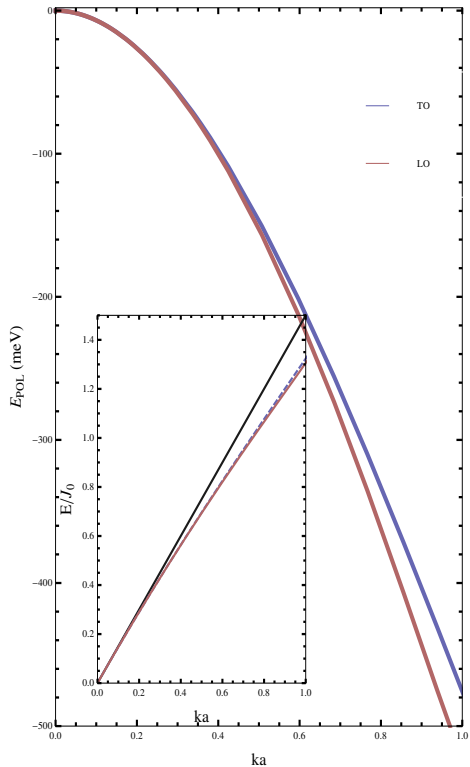}
\caption{(Color online) The polaron ground-state energy obtained from Eq.(%
\protect\ref{7}) as a function of $\overline{k}_{0}\boldsymbol{=}ka$. Inset:
shows the electron dispersion normalized to $J_{0}$ as a function of $%
\overline{k}_{0}\boldsymbol{=}ka$. }
\label{FIG1}
\end{figure}
\begin{figure}[bp]
\includegraphics [ height=11.27cm,width=7.cm]{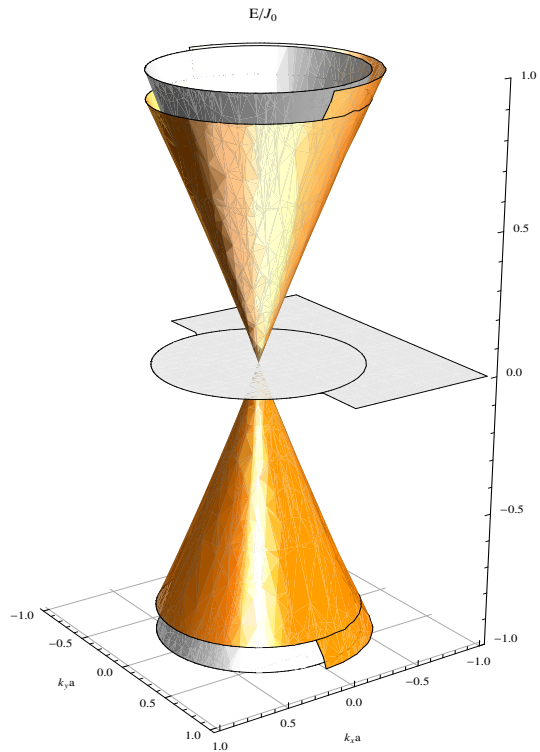}
\caption{Electron-hole band dispersions as a function of $\overline{k}_{0}%
\boldsymbol{=}ka$ both in the absence of (inner cone) and in the presence of
electron (hole)- phonon interaction (outer cone). Only LO phonon
contribution is taken into account. The thick dark circle on the $E/J_{0}=0$
plane is the projection of both perturbed (lower contour in the upper cone)
and unperturbed (upper contour in the upper cone) the contour at $E/J_{0}=1$%
. It is drawn to guide to eye how the dressed polaronic band dispersions
differ from the undressed ones.}
\label{FIG2}
\end{figure}
\begin{equation}
\boldsymbol{\Sigma }_{\lambda \lambda ^{^{\prime }}}^{0}(\mu ,k)=\frac{3%
\sqrt{3}}{4\pi r_{\mu }}J_{0}\alpha \left( 0\right) \ln \left[ \frac{1}{%
1-r_{\mu }\left( 2\overline{k}_{0}\right) ^{2}}\right] ,  \label{8}
\end{equation}%
where we have defined $\alpha \left( 0\right) =\left\vert \overline{\mathrm{M%
}}\right\vert ^{2}/4J_{0}\hbar \omega _{\Gamma }(0)$ and new dimensionless
wave vector, $\overline{k}_{o}=ka$. $\alpha \left( 0\right) $ is of order
of $0.054\left( 0.089\right) $ depending on the choice of $q_{0}=2\left(
2.5\right) $ $\mathring{A}^{-1}$.We analyze Eq.(\ref{7}) in FIG.~\ref{FIG1},
which shows both polaron self energy and Fermi velocity renormalization. As
it can be easily seen from the figure that polaron self energy is strongly
enhanced with increase in $\overline{k}_{o}$. This can also be justified
from its inset that shows how electron-phonon interactions renormalize the
Dirac velocity. To see this clearly we also present in FIG.~\ref{FIG2}
electron-hole band dispersions as a function of $\overline{k}_{0}\boldsymbol{%
=}ka$ both in the absence and presence of electron (hole)- phonon
interactions.

It is convenient to rewrite Eq.(\ref{8}) for small $k$'s, which corresponds
to neglect  phonon dispersions, i.e., it results with polaron
dispersion independent on $r_{\mu }$. To do this, we first expand the
logarithm in Eq.(\ref{8}), in power series of $k$ , and then replace the
resultant back into Eq.(\ref{8})so that Eq.(\ref{7}) reduces to the simple
form

\begin{equation}
\overline{E}_{\pm }=\pm \overline{k}_{o}\left\{ 1+4\left[ \frac{2\sqrt{3}}{%
\pi }\alpha \left( 0\right) \overline{k}_{0}\right] ^{2}\right\} ^{1/2}-%
\frac{2\sqrt{3}}{\pi }\alpha \left( 0\right) \overline{k}_{0}^{2},  \label{9}
\end{equation}%
where $\overline{E}_{\pm }=aE_{\pm }/\hbar v_{F}$ is the dimensionless
energy. The standard electron-self energy calculations due to the
interactions of electron with degenerate optical phonon modes with E$_{2g}$
symmetry in graphene predict a lowering of both conduction and valance band
energies in the same direction to preserve the symmetry of the Dirac cones.
Whereas, in our case, as $\overline{k}_{o}$ increases, negativity due to
intraband interactions is compensated by \ the second term in the
parenthesis in Eq.(\ref{8}), i.e., by the interband interactions. Although
it is compensated, their combined effects strongly modify the Dirac cones
(FIG. 2 ). Since Eq.(\ref{7}) can easily be rearranged into the form $%
\overline{E}_{\lambda }=\lambda \hbar \widetilde{v}_{F}\left( \overline{k}%
_{0}\right) \overline{k}_{0}$ , we can thus define the renormalized Fermi
velocity as

\[
\widetilde{v}_{F}\left( \overline{k}_{0}\right) =v_{F}\left[ \sqrt{1+4%
\widetilde{\alpha }^{2}\left( \overline{k}_{0}\right) }\mp \widetilde{\alpha
}\left( \overline{k}_{0}\right) \right] ,
\]%
wherein, in analogy with quantum chromodynamics\cite{GRIFFITHS1987}, we also
define a new coupling constant, running coupling constant, as a function of $%
k$, i.e, energy,

\[
\widetilde{\alpha }\left( \overline{k}_{0}\right) =\frac{2\sqrt{3}}{\pi }%
\alpha \left( 0\right)  \overline{k}_{0} .
\]

In conclusion, we predict a new type of polaronic formation in pristine
graphene. We demonstrate that our results are differ from those obtained
through standard electron self energy calculations due to electron-E$_{2g}$
phonon interactions in nondegenerate band case. We show that chiral polaron
band dispersions consist of k dependent terms besides the free undressed
one. In addition to free undressed ones, both intraband and interband
interactions coexist in a one polaron dispersion. Moreover, a considerable
renormalization of Fermi velocity, i.e., slope of Dirac cones in graphene is
observed due to the electron-optical phonon interactions in graphene. It is
also found that, the effect of LO phonon-electron interaction is stronger
than that of TO. As for the validity of our approximation it is valid for $%
\overline{k}_{0}\boldsymbol{=}ka<1,$ however it can be extended to higher $k$
values by just including the effect of trigonal warping in the total
Hamiltonian. It can also be extended to the case electron-$\mathbf{K}$
phonons interaction where a gap occurs. Furthermore, it can also be
generalized to the calculation of phonon induced electron-hole or
electron-electron interactions, i.e., to exciton or bipolaron binding
energies in graphene.

\begin{acknowledgments}
I thank Professor T. Altanhan for valuable discussions.
\end{acknowledgments}

\end{document}